%% file: eqpp.tex
\documentclass[12pt,a4paper]{article} 
\usepackage{amsmath, amsfonts, url,amsthm} 
\usepackage{graphicx} 
\usepackage{pgf,tikz}
\usepackage{natbib}

\title{ $P$-values, $q$-values and posterior probabilities for
  equivalence in genomics studies}
\author{J. Tuke\footnote{To whom correspondence should be addressed}\\
{\it School of Mathematical Sciences}\\ 
{\it The University of Adelaide, Adelaide SA 5005, Australia}\\ simon.tuke@adelaide.edu.au\\\\
G.F.V. Glonek\\ 
{\it School of Mathematical Sciences}\\
{\it The University of Adelaide, Adelaide SA 5005, Australia}\\ gary.glonek@adelaide.edu.au\\\\ 
P.J. Solomon\\
{\it School of Mathematical Sciences}\\
{\it The University of Adelaide, Adelaide SA 5005, Australia}\\
patty.solomon@adelaide.edu.au}

\date{}

\begin{document}

\maketitle

\begin{abstract}
  Equivalence testing is of emerging importance in genomics studies
  but has hitherto been little studied in this content.  In this
  paper, we define the notion of equivalence of gene expression and
  determine a \lq strength of evidence' measure for gene equivalence.
  It is common practice in genome-wide studies to rank genes according
  to observed gene-specific $P$-values or adjusted $P$-values, which
  are assumed to measure the strength of evidence against the null
  hypothesis of no differential gene expression.  We show here, both
  empirically and formally, that the equivalence $P$-value does not
  satisfy the basic consistency requirements for a valid strength of
  evidence measure for equivalence. This means that the widely-used
  $q$-value \citep{Storey:2002} defined for each gene to be the
  minimum positive false discovery rate that would result in the
  inclusion of the corresponding $P$-value in the discovery set,
  cannot be translated to the equivalence testing framework.  However,
  when represented as a posterior probability, we find that the
  $q$-value does satisfy some basic consistency requirements needed to
  be a credible measure of evidence for equivalence.  We propose a
  simple estimate for the $q$-value from posterior probabilities of
  equivalence, and analyse data from a mouse stem cell microarray
  experiment which demonstrate the theory and methods presented here.
\end{abstract}



\section{Introduction}

\subsection{Motivation for equivalence testing in genomics}

Equivalence testing is pervasive in the pharmaceutical industry for
assessing whether two drugs or treatment regimens provide comparable
therapeutic effects within pre-defined clinical and statistical limits
\citep{Senn:2001,Chow:2008}.  Typically, a new treatment formulation
is of interest if it potentially provides an equivalent therapeutic
effect with fewer adverse side-effects than the standard treatment, or
is less costly to produce.  In such trials, a fixed level of
statistical significance is usually prescribed and within this
context, tests based on confidence interval inclusion provide a sound
basis for analysis \citep{Westlake:1972,Wellek:2003}.  \medskip

Recently, a number of important applications entailing hypotheses of
equivalence have arisen in the area of microarray bioinformatics and
genomics studies, as illustrated by the following.
\medskip

\noindent \emph{Illustration 1: Gene profiling}.  There are biological
problems in which it is desired to establish that the levels of
expression for certain genes remain constant across different
conditions and/or points in time. \cite{Tuke:2009} propose a general
method for ranking genes according to their conformance to a
pre-specified profile of expression over time, known as {gene
  profiling}.  In many situations, the expected time-course profile
dictates that gene expression should remain the same, or be {\it
  equivalent}, at two or more different time points, and hence
equivalence testing methods are required as part of the statistical
inferential framework.  \medskip

\noindent \emph{Illustration 2: Experimental study of regulatory $T$
  cells of the immune system}. $CD4^+CD25^+$ regulatory $T$~cells
(known as $CD4^+$ Treg cells) play a central role in the human immune
system, including in the immunopathogenesis of autoimmune diseases,
tumours, viral infections such as HIV, and organ transplants
\citep{Wei:2006}.  Naturally occurring $CD4^+$ Treg cells are
available only in very small quantities however, and for this reason
many \emph{in vitro} experiments are conducted on induced cell
populations. To ensure the integrity of such experiments, it is
desirable to establish that gene expression is equivalent between the
naturally occurring and induced cells, at least under baseline
conditions. The establishment of such conditions requires direct
formal testing of an equivalence hypothesis, rather than simply
failing to find evidence of differential expression between the two
populations.  \medskip

\noindent \emph{Illustration 3: Normalisation and quality assurance}.
A third and common problem for which methods of equivalence testing
are appropriate in practice (although not always applied) is in the
identification of highly stable genes or set of genes for the purpose
of quality assurance, or for normalisation for use in future
experiments. The identification of such genes (often housekeeping
genes of some variety) is particularly important in large-scale
studies utilising a reference population and conducted over time, such
as in cancer studies. The reference population may be subject to
temporal or other changes, and linkage experiments are usually
conducted to ensure comparability of results and diagnoses over time.
\bigskip

In each of these illustrations, the aim of the scientific experiment
is to test one or more hypotheses of equivalence.  In a typical
experiment, many thousands of genes will be tested simultaneously, and
a primary goal of an initial bioinformatics exploratory analysis,
known as a \lq gene-screen', is to produce a ranked gene list.
Therefore, we require an inferential strength of evidence measure for
equivalence which will give just such a ranking.

Now, we know that there are two types of mistakes we can make when
testing a statistical hypothesis: the first is Type I error, which is
the probability of rejecting the null hypothesis when it is true, and
the second is Type II error, which is the probability of retaining the
null hypothesis when it is false.  Type I errors are also known as
false positives, and it is their dramatically increased frequency in
genomics applications which has received most attention.  A popular
approach to adjusting for multiple hypotheses testing in this context
is to find the observed $P$-value for each gene, and then to calculate
its associated $q$-value \citep{Storey:2003}.  For each gene, the
observed $q$-value is the positive false discovery rate (pFDR) that would result
in the corresponding $P$-value being included in the discovery set,
should this be the adjustment procedure applied.  The pFDR itself is
the rate at which genes are incorrectly discovered to be significant;
there are also other methods for controlling the Type I error at a
reasonable level, for example, controlling the overall family-wise
error rate, but the pFDR is popularly applied in genomics studies.

\medskip

In Section 2, we review the formal definition of the $q$-value, then
derive the equivalence $P$-value for each gene. We show that our
$P$-value equates to alternative derivations by \cite{Senn:2001} and
\cite{Ge:2003}. We then point out problems with equivalence
$P$-values, in particular, that they tend to zero as the standard
error of the estimator increases, and that they are non-monotonic,
thus rendering them unsuitable as a basis for calculating
equivalence $q$-values.  In Section 3, we describe the $q$-value as a
posterior probability, and prove that the equivalence $q$-value
obtained in this way is a monotonic function of the standard error of
the estimator of interest, thus satisfying the basic consistency
requirement (among other things) for a strength of evidence measure
for the gene ranking. We also show how to estimate $q$-values from the
observed posterior probabilities for equivalence.

In Section 4, we set out our motivating analysis of a murine stem cell 
microarray experiment conducted at the University of Adelaide. The overall aim 
of the larger time course experiment was to identify and investigate genes 
involved in pluripotency. We are interested in which of the $23, 040$ mouse 
genes are equivalently expressed at day 0 ({\it i.e.}, the beginning of the 
experiment) and at day 3. We begin by proposing a joint probability model for 
the observed and true log ratios to obtain the posterior probability of 
equivalence; we employ an empirical rather than a fully Bayesian approach to 
estimation, utilising the EM algorithm to estimate the specified 
hyperparameters. We finish with a brief conclusion in Section~\ref{sec:con} 
extolling the virtues of the $q$-value, when represented as a posterior 
probability, as a credible measure of evidence for equivalence.

\section{Equivalence $P$-values and $q$-values}

\subsection{The positive false discovery rate and the $q$-value}
\label{sec:qv}

We are interested in the positive false discovery rate (pFDR) due to
\cite{Storey:2003}. To review, Table~\ref{tab:Storey} gives the
possible outcomes when $m$ hypotheses tests are performed according to
some significance rule:

\begin{table}[htbp]
\begin{center}
\begin{tabular}{lccc}
\hline
&Accept null & Reject null & Total\\
\hline
Null true & U & V & $m_0$\\
Alternative true & T & S & $m_1$\\
\hline
&W & R & m\\
\hline
\end{tabular}
\caption{Possible outcomes from $m$ hypotheses tests.}
\label{tab:Storey}
\end{center}
\end{table}
The positive false discovery rate  is then defined  as
\[
	\text{pFDR}=E\left[V/R|R>0\right] 
\]


In genomics studies, and for equivalence testing in particular, $R>0$
almost always, so we assume from now on that the pFDR is the same as
the FDR.

Now suppose that for each of the  $m$ hypotheses, the test statistics \\
$T_1,T_2,\ldots,T_m$ are observed. Consider a nested set of
significance regions denoted by $\Gamma_\alpha, 0\leq \alpha\leq1$,
where $\alpha$ is such that
$$
P(T_i\in\Gamma_\alpha|i\text{th null hypothesis is true})=\alpha
$$
and $\alpha'\leq\alpha$ implies that
$\Gamma_{\alpha'}\subset\Gamma_{\alpha}$.  
\cite{Storey:2003} defines
the $q$-value for an observed test statistic $T=t$ to be
\begin{eqnarray}
	q\text{-value}(t)=\inf_{\Gamma_\alpha:t\in \Gamma_\alpha}
	\text{pFDR}(\Gamma_\alpha)
\end{eqnarray}

We return to the general definition (1) in Section~\ref{sec:qv_prop}.

\subsection{$P$-values for equivalence testing}


Consider a parameter of interest $\theta$ and
its associated estimator, $\hat{\theta}$, such that
\[
	\hat{\theta}\sim N(\theta,SE(\hat{\theta})^2),
\] 
{\it i.e.}, $\hat \theta$ is an unbiased estimator of $\theta$. For
simplicity, but without loss of generality, we assume that the
standard error of the estimator $\hat \theta$ is known.

In statistical equivalence testing, the null and alternative
hypotheses are, respectively,
\begin{align}
	H_0&:|\theta|\geq \varepsilon, \quad \varepsilon>0,\label{eq:H1}\\
	H_A&:|\theta|<\varepsilon\label{eq:H2}
\end{align}

Consider now the test statistic,
\begin{equation}
	U(\hat{\theta})=\frac{\varepsilon-|\hat{\theta}|}
	{SE(\hat{\theta})}\label{eq:U}
\end{equation}
The test statistic is chosen so that large values give evidence in
favour of $H_A$.  From \eqref{eq:U}, we can deduce that as the
observed value of the estimator gets closer to zero from either
direction, then the test statistic increases to a maximum at
$\varepsilon/SE(\hat \theta)$, while as the observed value of the
estimator moves away from zero, the test statistic decreases.
\medskip

 The following theorem is from Casella (2002,  p.397)\nocite{Casella:2002}:
 \medskip

 \noindent {\bf Theorem 1}: Let $W(\boldsymbol{X})$ be a test
 statistic such that large values of $W$ give evidence that $H_A$ is
 true. For each sample point $\boldsymbol{x}$, define
\begin{equation*}
		p(\boldsymbol{x})=\sup_{\theta\in\Theta_0}
		P_{\theta}(W(\boldsymbol{X})\geq W(\boldsymbol{x})),
\end{equation*}
where $\Theta_0$ is a subset of the parameter space defined by the
null hypothesis.  Then, $p(\boldsymbol{X})$ is a valid $P$-value.

\bigskip

Applying Theorem 1 to the test statistic \eqref{eq:U} yields the
$P$-value
\begin{align}
	P_U&=\sup_{\theta\in \Theta_0}
	P_{\theta}(U(\hat{\theta})\geq u(\hat{\theta}))\cr
	&=\sup_{|\theta|\geq\varepsilon}P_{\theta}
	\left(
		\frac{-|\hat{\theta}|-\theta}{SE(\hat{\theta})}
		\leq Z \leq  
		\frac{|\hat{\theta}|-\theta}{SE(\hat{\theta})}
	\right),\label{eq:PVZ}
\end{align}
where $\hat{\theta}$ is the observed estimate and $Z\sim
N(0,1)$. It can be shown that $P_U$ is maximised at
$\theta=\pm\varepsilon$, so that
\begin{align}
	P_U = P\left(
	\frac{-|\hat{\theta}|-\varepsilon}{SE(\hat{\theta})}
	\leq Z \leq  
	\frac{|\hat{\theta}|-\varepsilon}{SE(\hat{\theta})}
	\right)
	\label{eq:EPV}
\end{align}

\bigskip

An alternative derivation of $P_U$ uses a Neyman-Pearson-type test, as
proposed by \cite{Senn:2001}.  He considers the statistical problem of
equivalence testing for a parameter of interest $ \theta$, with an
estimator $\hat \theta$, such that
\[
	\hat{\theta}\sim N(\theta,SE(\hat{\theta})^2)
\]
Then for a pair of symmetric critical values for $\hat \theta$, $-c$
and $c$, the power function for a test based on $\hat \theta$ is
\begin{eqnarray}
     \Phi\left(\frac{c-\varepsilon}{SE(\hat \theta)}\right)-
     \Phi\left(\frac{-c-\varepsilon}{SE(\hat \theta)}\right)\label{eq:PF}
\end{eqnarray}
To achieve a test of significance level $\alpha$, equation
\eqref{eq:PF} is set equal to $\alpha$ and solved for $c$.

In other work, \cite{Ge:2003} state that the $P$-value is the minimum
Type 1 error over all possible rejection regions that contain the
observed value of the test statistic. Again consider the observed
estimate $\hat \theta$. The $P$-value associated with this observed
estimate, which is denoted by $P_{alt}$ can therefore be derived from
the power function \eqref{eq:PF}:
\begin{align*}
	P_{alt}&=\min_{c:\hat \theta\in(-c,c)}
	\Phi\left(\frac{c-\varepsilon}{SE(\hat \theta)}\right)-
     \Phi\left(\frac{-c-\varepsilon}{SE(\hat \theta)}\right)\\
	&=\Phi\left(\frac{|\hat \theta|-\varepsilon}{SE(\hat \theta)}\right)-
     \Phi\left(\frac{-|\hat \theta|-\varepsilon}{SE(\hat \theta)}\right)\\
	&=P_U
\end{align*}
which again recovers our estimate (6).


\subsection{Inconsistency of equivalence $P$-values}

The $P$-value is often interpreted as a strength of evidence measure
for the alternative hypothesis. For example, \cite{Casella:2002} state
that small values of the $P$-value give evidence that the alternative
hypothesis is true (Definition 8.3.26, page 397). This is not true,
however, for equivalence $P$-values, as we now demonstrate.  \medskip

\begin{figure}[htbp]
	\centering
		\includegraphics[width=3.5in]{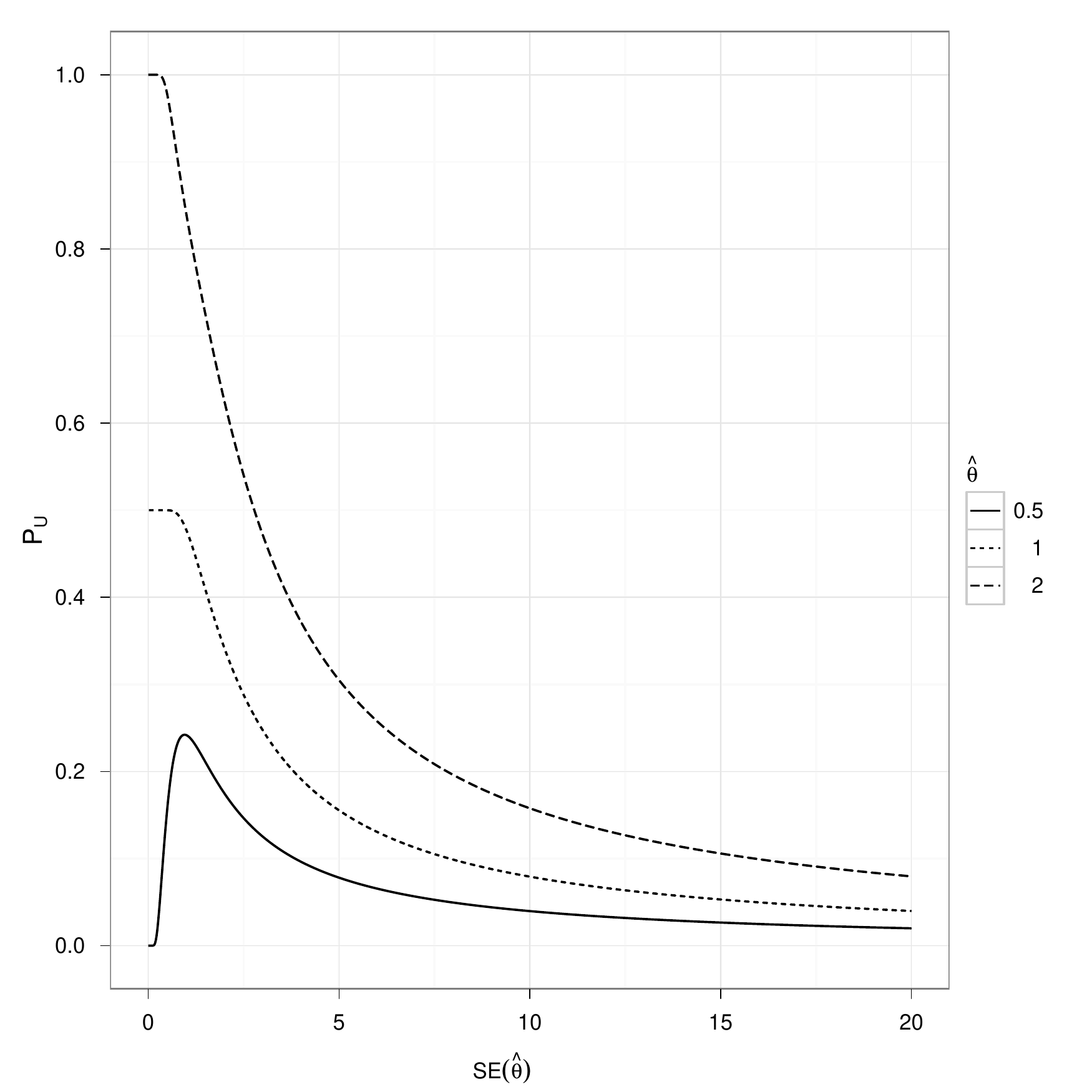}
                \caption{Equivalence $P$-value versus $SE(\hat
                  \theta)$ for an equivalence margin of
                  $\varepsilon=1$.}
	\label{fig:figures_eqpv}
\end{figure}

Consider Equation $(6)$ as a function of $SE(\hat \theta)$, with $\hat
\theta$ and $\varepsilon$ fixed. In general, as $SE(\hat \theta)$
increases, then $P_U$ decreases to zero. This is illustrated in
Figure~\ref{fig:figures_eqpv} which plots $P_U$ versus $SE(\hat
\theta)$ for values of $\hat \theta$ of $0.5$, $1$, and $2$, and
$\varepsilon=1$. For each value of $\hat \theta$, as $SE(\hat \theta)$
increases, then $P_U$ decreases to zero. Thus, small values of $P_U$
do not necessarily indicate evidence that the alternative hypothesis
is true, {\it i.e.}, that $\theta$ is equivalent to zero.  For
example, consider an observed value of $\hat \theta=0.5$, with
$SE(\hat \theta)=10$. For these observed values, we are $95$\%
confident that the true value of $\theta$ lies between $-19.09964$ and
$20.09964$, but the observed value of $P_U$ for the equivalence margin
of $(-1,1)$ is $0.03969$.  This small value of $P_U$ indicates strong
evidence that $\theta$ lies between $-1$ and $1$, which is false.  In
fact, the largest confidence interval that would be wholly contained
within the equivalence margin is a $52$\% confidence interval.

A further observation of note is the lack of monotonicity of $P_U$ as
$SE(\hat \theta)$ increases.  This is illustrated by the graph of
$P_U$ versus $SE(\hat \theta)$ for the case $\hat \theta=0.5$ in
Figure~\ref{fig:figures_eqpv}.  As $SE(\hat \theta)$ increases from
$0$ to $1$, $P_U$ increases to a maximum of about $0.24$, then $P_U$
decreases as $SE(\hat\theta)$ increases from $1$ to $20$.  As a result
of this lack of monotonicity, it is possible to have the same value of
$P_U$ for different values of $SE(\hat\theta)$. For example, the $P_U$
for an observed value of $\hat \theta=0.5$, an equivalence margin of
$\varepsilon$ equal to $1$, and $SE(\hat \theta)=0.3$, is
$0.04779$. This is the same $P$-value that is observed for the same
conditions but with $SE(\hat \theta)$ equal to $8.28224$. To assign
the same strength of evidence to both of these cases is incorrect.

\medskip

These inconsistencies in interpretation of the observed equivalence
$P$-values highlight the fact that they lack the basic consistency
requirements necessary to provide a valid strength of evidence measure
for equivalence, either on their own as raw (unadjusted) $P$-values,
or as a basis for (adjusted) statistics such as the $q$-value.
\medskip

We turn instead to a Bayesian formulation, and in particular, propose
the posterior probability of equivalence, as an alternative strength
of evidence measure for equivalence.  We justify this approach in the
next section and also show how to calculate approximate $q$-values
from the posterior probabilities.

\section{Monotonicity of posterior probabilities for equivalence and
  $q$-values}
\label{sec:qv_prop}

In Section~\ref{sec:qv}, we defined the $q$-value, $q(t)$, for $m$ identical
hypothesis tests with corresponding test statistics $T_1,\dots, T_m$
and significance region $\Gamma$ \citep{Storey:2003}.

For each hypothesis test, there is also a corresponding Bernoulli
random variable $H_i$ with $H_i=0$ if the null hypothesis is true and
$H_i=1$ if the null hypothesis is false. \cite{Storey:2003} assumes
that $(T_i,H_i)$ are independent identically distributed random
variables, $T_i|H_i\sim (1-H_i)F_0+H_iF_1$ for some null distribution
$F_0$ and alternative distribution $F_1$, and
$H_i\sim$Bernoulli$(\pi)$ for $i=1,\ldots,m$, where $\pi$ is the
\emph{a priori} probability that a null hypothesis is false, {\it i.e.},
$P(H_i=1)=\pi$.  Therefore, the $q$-value is the infimum of the
quantity $P(H=0|T\in \Gamma_\alpha)$, that is, the posterior
probability that the null hypothesis is true given that the test
statistic is contained in the significance region of level $\alpha$.
\medskip

The following theorem states that for equivalence testing, the
$q$-value is monotonically decreasing for increasing variance of the
test statistic.  We state the theorem here and defer its proof to
Appendix 2, which also contains the statement and proof of a lemma
which is used in the proof of Theorem 2.


\medskip

\noindent {\bf Theorem 2}: Suppose \(\theta\) has prior distribution
\(P(\theta)\) and \(T|\theta\sim N(\theta,\sigma^2)\). Consider
numbers \(0<\ell<\varepsilon\) and assume further that
\(0<P(-\varepsilon<\theta<\varepsilon)<1\).  Then
\(P(-\varepsilon<\theta<\varepsilon\mid-\ell<T<\ell)\) is a decreasing
function of \(\sigma^2\).  \bigskip

Note that for given $SE({\hat\theta})$, the $q$-value is intended to
estimate the posterior probability
\[
1 -	P(-\varepsilon<\theta<\varepsilon\mid-\ell<T<\ell),
\]
which we have therefore established increases monotonically with
$SE({\hat \theta})$.

\subsection{Estimating $q$-values from posterior probabilities of equivalence}

Before illustrating the methods on the stem cell microarray data, we
take the development a step further and show how to estimate
$q$-values from the observed posterior probabilities of equivalence in
any given application.

Consider an arbitrary cutoff point $t$ such that all genes will be
considered equivalent if their observed posterior probability of
equivalence is greater than or equal to $t$. The FDR for the cutoff
$t$ is then
\[
	\text{FDR}(t)=E\left[\frac{V(t)}{R(t)}\right],
\]
where $R(t)$ is the number of genes considered equivalent with assumed
cutoff $t$, and $V(t)$ is the number of genes considered equivalent
with cutoff $t$ that are not in fact equivalent, {\it i.e.}, these are
false positives.  \cite{Storey:2003} recovers the general result that
\[
	E\left[\frac{V(t)}{R(t)}\right]\approx \frac{E[V(t)]}{E[R(t)]}
\]
In the case of gene equivalence studies, $E[R(t)]$ can be estimated by
the number of genes whose posterior probability of equivalence is
greater than or equal to $t$. The estimate of $E[V(t)]$ is obtained by
considering each hypothesis test as a Bernoulli random variable
$X_i,i=1,\ldots,m$, where $X_i=1$ if the null hypothesis is true, that
is, $|\theta_i|\geq \varepsilon$, and $X_i=0$ if the null hypothesis
is false. Therefore,
\[
	 E[V(t)]= E\left[\sum_{i:p_i\geq t}X_i\right]=\sum_{i:p_i\geq t} E[X_i]
	=\sum_{i:p_i\geq t}q_i,
\]
where
\[
	q_i = P(|\theta_i|\geq\varepsilon),
\]
and $p_i$ is the posterior probability of equivalence for the $i^{th}$
gene.

The value of $q_i$ is unknown but can be estimated from the posterior
distribution of $\theta_i$ by $q_i\approx 1-p_i$, where $p_i$ is the
posterior probability of equivalence of the $i$th gene. The estimated
$q$-value for a cutoff point $t$ is then
\begin{align}
	\hat q(t) &=
	\frac{\sum_{i:p_{i}\geq t}(1-p_{i})} 
	{\#\{i:p_{i}\geq t  \}},\label{eq:qv}	
\end{align}
where $\#A$ represents the cardinality of the set $A$.
\medskip

In the next section we explore the performance of the posterior
probabilities of equivalence for the stem cell data.  We also estimate
the approximate $q$-values obtained from the posterior probabilities
of gene expression equivalence.

\section{Gene equivalence in mouse stem cells on day 0 and day 3}

Much of our work on gene equivalence has been motivated by a stem cell
microarray experiment conducted at the University of Adelaide. The
overall aim of the experiment was to identify and investigate genes
involved in the cellular process of pluripotency; the details of this
study and its design are described in \cite{Tuke:2009} and
\cite{Tuke:2012}.  Here we are interested in which of $23, 040$ mouse
genes are equivalently expressed at day 0 ({\it i.e.}, the beginning
of the experiment) and at day 3.  This equivalence gene-set will
include genes which are not expressed on either day and genes which
are expressed on both day 0 and day 3.  We know {\it a priori} that
there are housekeeping genes included on the microarray which are
highly and constantly expressed across hybridisations, and that there
are genes involved in pluripotency which are also expressed on both
days.  There are two (dye-swapped) two-colour mouse Compugen
long-oligonucleotide microarrays available for analysis, which are
treated as independent replicates.

We proceed as follows: to begin, we specify a joint distribution for
the true mean log ratio and the observed mean log ratio for each gene,
in which the prior distribution for the true mean log ratio is a
mixture model of three normal distributions. The hyperparameters for
the joint distribution are then estimated from the observed mean log
ratios using the EM algorithm (as described in Appendix 1) and finally
the posterior distribution of the true gene log ratio is derived and
used to calculate the posterior probability of equivalence.  We do not
use a fully Bayesian formulation for our case study, but rather
estimate the hyperparameters then treat them as \lq known', thereby
utilising a simpler empirical Bayesian approach.

\subsection{The probability model for each gene}

The parameter of interest is the true mean log ratio which will be
equal to zero if the gene is equivalently expressed on day 3 and day
0.

If the true mean log ratio for gene $ i$ is denoted by ${\theta_i}$,
$i = 1,...,23040$, then the sample mean of the observed log ratio,
$Y_i$, is assumed to be normally distributed with mean $\theta_i$ and
variance $\sigma^2_i$, which is assumed known. The probability density
function for $Y_i$ is then
\[
f(y_i|\theta_i)=\frac{1}{\sqrt{2\pi\sigma_i^2}}\exp\left[-\frac{(y_i-\theta_i)^2}{2\sigma_i^2}\right]
\]

\begin{figure}[htbp]
	\centering
        \includegraphics[width=3.5in]{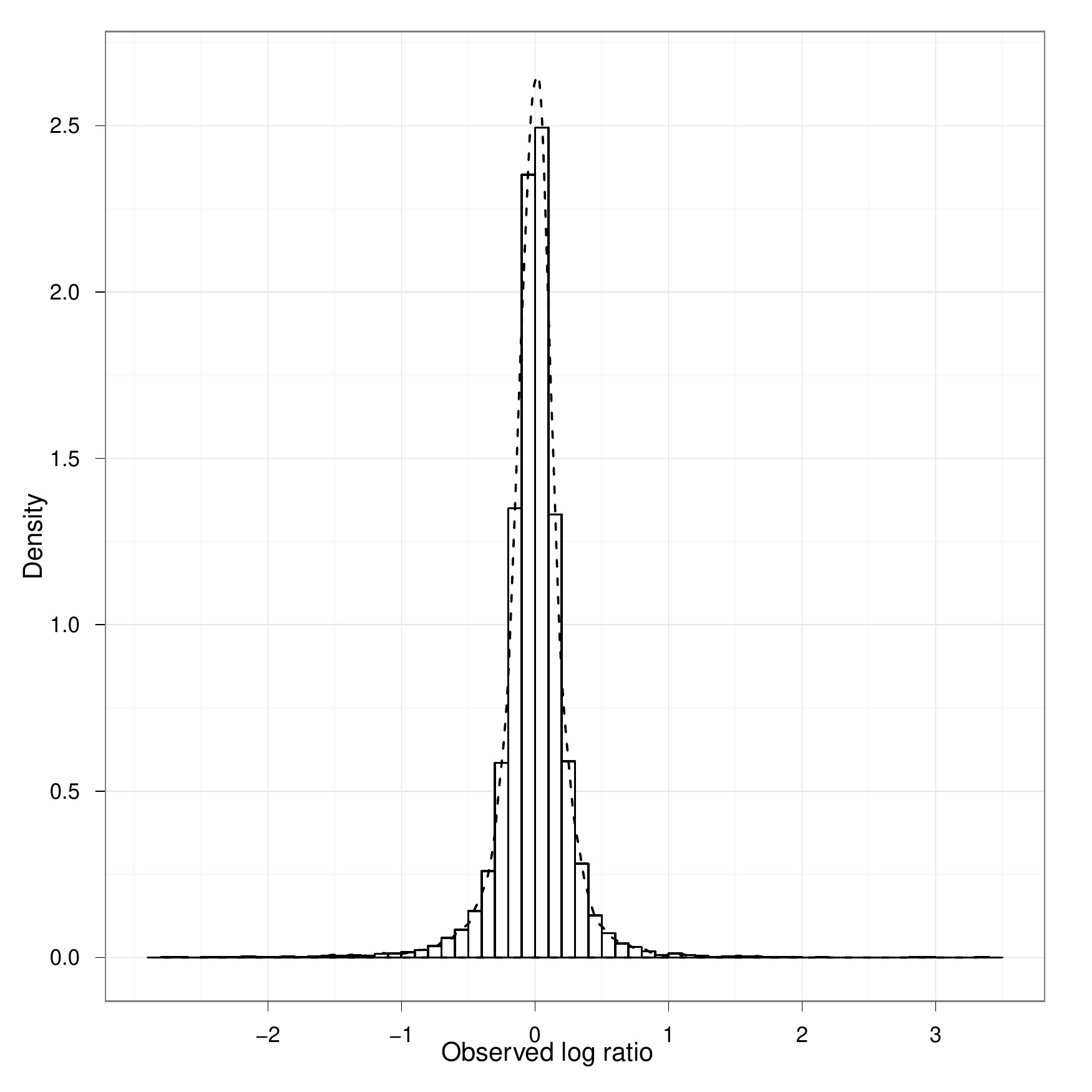}
	\caption{Histogram of  observed mean log ratios for day 3 compared to day 0 for 
	the mouse stem cell data.}
	\label{fig:LRd3d0}
\end{figure}

The prior probability distribution for $\theta_i$ is assumed to be a
mixture model of three normal distributions. This choice of prior
gives flexibility to the full probability model and is motivated by
the data.  Figure~\ref{fig:LRd3d0} presents a histogram of the
observed mean log ratios for day 3 compared to day 0 for the data
$y_i$, and shows a symmetric bell-shaped density with relatively heavy
tails compared to a normal density.  This suggests a $t$-distribution
with a small number of degrees of freedom would be an appropriate
prior, among other possibilities, and it is straightforward to show
that such a $t$-distribution is well approximated by a mixture of
three normal distributions.


The prior distribution for $\theta_i, i=1,\ldots,23040$ is therefore
\[
f(\theta_i)=\sum^3_{j=1}\frac{\pi_j}{\sqrt{2\pi\tau^2_j}}\exp\left[-\frac{(\theta_i-\mu_j)^2}{2\tau^2_j}\right]
\]
with hyperparameters $\pi_j,\mu_j,\tau^2_j$, $j=1,2,3$ such that
$\sum_{j=1}^3\pi_j=1$.

The full probability model for each gene $i$ is then
\begin{align}
  f(y_i,\theta_i)&=f(y_i|\theta_i)f(\theta_i)\cr
  &=\frac{1}{\sqrt{2\pi\sigma_i^2}}\exp\left[-\frac{(y_i-\theta_i)^2}{2\sigma_i^2}\right]\sum^3_{j=1}\frac{\pi_j}{\sqrt{2\pi\tau^2_j}}\exp\left[-\frac{(\theta_i-\mu_j)^2}{2\tau^2_j}\right]\label{eq:BM}
\end{align}

Under this model, the posterior density function of the true mean log
ratio for the $i$th gene, $\theta_i$, $ i=1,\ldots,23040$, is
\begin{align}
	f(\theta_i| y_i)&=\frac{f(y_i|\theta_i)f(\theta_i)}{f(y_i)}\notag\\
					&=\frac{\frac{1}{\sqrt{2 \pi \sigma_i^2 }}
					\exp\left[-\frac{(y_i-\theta_i)^2}{2\sigma_i^2}\right]
					\sum^3_{j=1}\frac{\pi_j}{\sqrt{2\pi\tau^2_j}}
					\exp\left[-\frac{(\theta_i- \mu_j)^2}
						{2 \tau^2_j }\right]}
					{\sum_{j=1}^3\frac{\pi_j}
					{\sqrt{2 \pi (\sigma_i^2 + \tau^2_j) }}
					\exp\left[-\frac{(y_i-\mu_j)^2}
					{2 (\sigma_i^2 + \tau^2_j) }\right]}
						\label{eq:PPE1}
\end{align}
Rearranging \eqref{eq:PPE1} gives
\begin{align}
	f(\theta_i | y_i ) = \frac1A \sum_{j=1}^3 B_jC_j
		\exp \left[\frac{E^2_j}{2D^2_j}\right]
		\sqrt{2\pi D^2_j}\frac1{\sqrt{2\pi D^2_j}}
		\exp \left[-\frac{(\theta_i-E_j)^2 }{2D^2_j}\right],\label{eq:PP}
\end{align}
where
\begin{align*}
	A	&=\sum_{j=1}^3\frac{\pi_j}{\sqrt{2 \pi \rho^2_j }}
			\exp\left[-\frac{(y_i-\mu_j)^2}{2 \rho^2_j }\right],\\
	B_j	&=\frac{\pi_j}{\sqrt{2 \pi \sigma_i^2 }\sqrt{2 \pi \tau^2_j }},\\
	C_j	&=\exp \left[-\frac{(\tau^2_j y_i^2+ \sigma_i^2 \mu_j^2) }
		{2 \sigma_i^2 \tau^2_j  }\right],\\
	D_j^2 &=\frac{\sigma_i^2 \tau^2_j  }{\tau^2_j + \sigma_i^2 }, 
	\text{ and}\\
	E_j &=\frac{y_i \tau^2_j + \mu_j \sigma_i^2 }{\tau^2_j + \sigma_i^2 }
\end{align*}
Explicit calculation of the normalising constant is a straightforward
adaptation of the standard argument used for a Gaussian mean with a
Gaussian prior; see for example, \cite{Gelman:2004}.

The posterior density \eqref{eq:PP} is used to calculate the
conditional probability that $\theta_i$ lies within the equivalence
region, conditional on the observed mean log ratio.  That is,
$P(-\varepsilon < \theta_i < \varepsilon | y_i)$ is equal to
\begin{eqnarray}
		 \frac1A \sum_{j=1}^3 B_jC_j
			\exp \left(\frac{E^2_j}{2D^2_j}\right)
			\sqrt{2\pi D^2_j}
			\int^\varepsilon_{-\varepsilon}
			\frac1{\sqrt{2\pi D^2_j}}
			\exp \left[-\frac{(\theta-E_j)^2 }{2D^2_j}\right] d\theta_i
			\label{eq:EqPP}
\end{eqnarray}

As already noted, this is not a fully Bayesian formulation since we
have not specified distributions for the hyperparameters $\pi_j,\mu_j$
and $\tau^2_j$.  Rather, we take an empirical Bayes approach to obtain
point estimates for each of the nine hyperparameters, which are then
substituted into the posterior distribution of $\theta_i$ to obtain
posterior probabilities of equivalence.  We employ the EM algorithm to
estimate the hyperparameters, and the details are given in Appendix 1.


\subsection{Application to the stem cell data: day 3 compared to day 0}

We know from Equation (11) that the posterior distribution of
$\theta_i$, $i = 1, . . . , 23040$, is a weighted mixture of the
likelihood of $\theta_i$ given the data $y_i$ and the prior
distribution of $\theta_i$, $i = 1, . . . , 23040$.  We study first
the behaviour of the posterior probability as a function of the
variance $\sigma_i^2$.

\begin{figure}[htbp]
  \centering
  \includegraphics[width=3.5in]{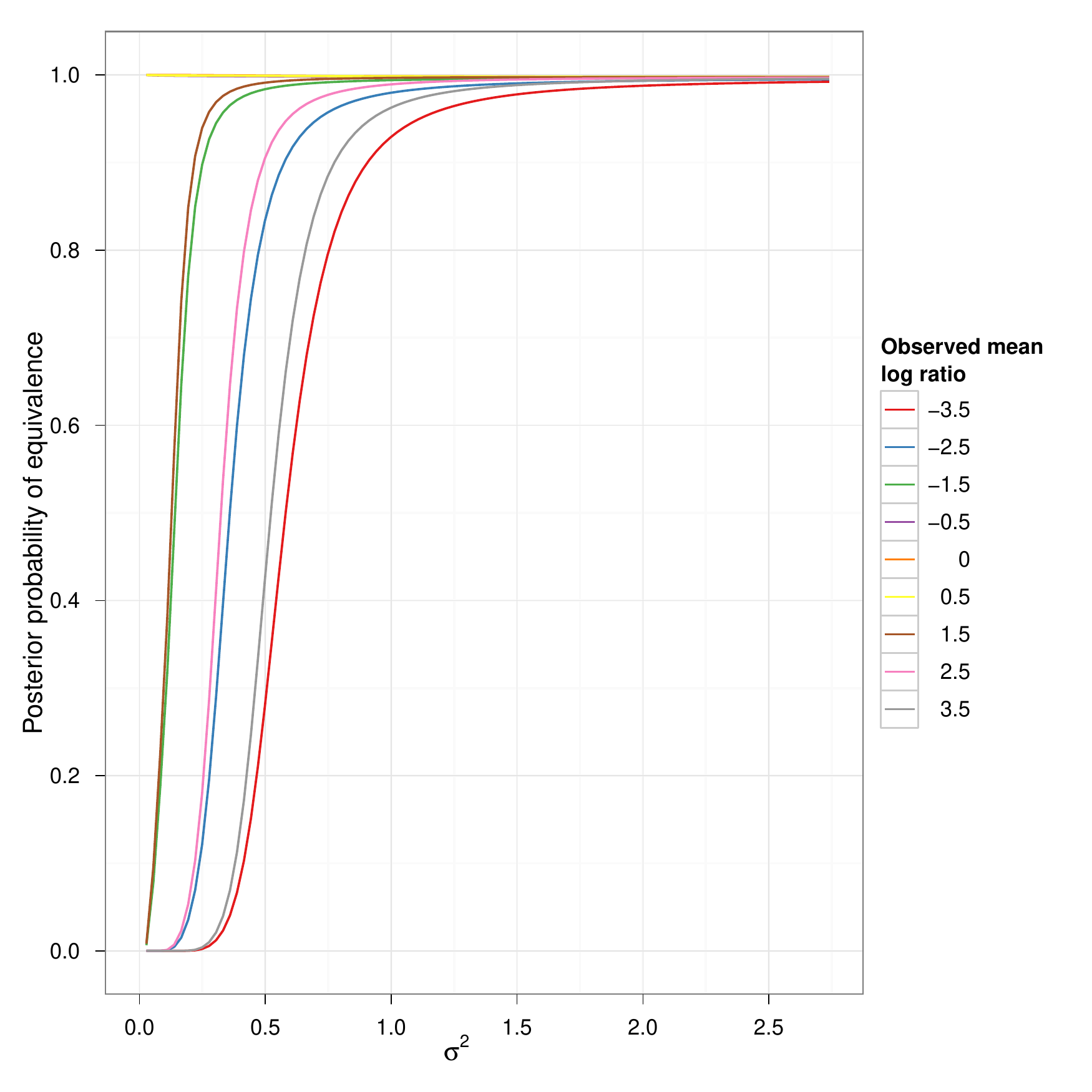}
  \caption{Posterior probability of equivalence versus variance of the
    mean log ratio, $\sigma_i^2$, for the given mean log ratio value.}
  \label{fig:EqPP1}
\end{figure}

Observe that if the variance of the mean log ratio $\sigma^2_i$ is
zero, then $y_i$ is equal to $\theta_i$ and the gene is equivalently
expressed if $|y_i| < \epsilon = 1$ and not equivalently expressed if
$|y_i|\geq 1$.  This is illustrated in Figure \ref{fig:EqPP1} which
plots the posterior probability of equivalence versus the variance of
the mean log ratio, $ \sigma^2_i$, for mean log ratios over a range of
given values.  The estimates of the hyperparameters used for the
calculations implicit in this Figure are given in
Table~\ref{tab:EM_res} of Appendix 1.

For mean log ratios contained within $(-1,1)$ the posterior
probability of equivalence for a value of $\sigma_i^2=0$ is one, while
those that lie outside $(-1,1)$ have a posterior probability of
equivalence of zero when $ \sigma_i^2$ is zero.  As the variance of
the mean log ratios increases, the posterior distribution of
$\theta_i$, $i=1,\ldots, 23040$, is weighted towards the prior
distribution of $ \theta_i$. This is demonstrated by the posterior
probability of equivalence increasing to one for all values of the
mean log ratio as the value of $ \sigma_i^2$ increases to $2.5$ in
Figure~\ref{fig:EqPP1}.  Of note is that the posterior probability of
equivalence is higher for the (positive) mean log ratios $1.5, 2.5$
and $3.5$ compared to the negative means $-1.5, -2.5$ and $-3.5$, for
the same value of $ \sigma_i^2$. This is because the empirical prior
distribution of $ \theta_i$, $i=1,\ldots,23040$, is asymmetric about
zero.

\begin{figure}[htbp]
	\centering
	\includegraphics[width=3.5in]{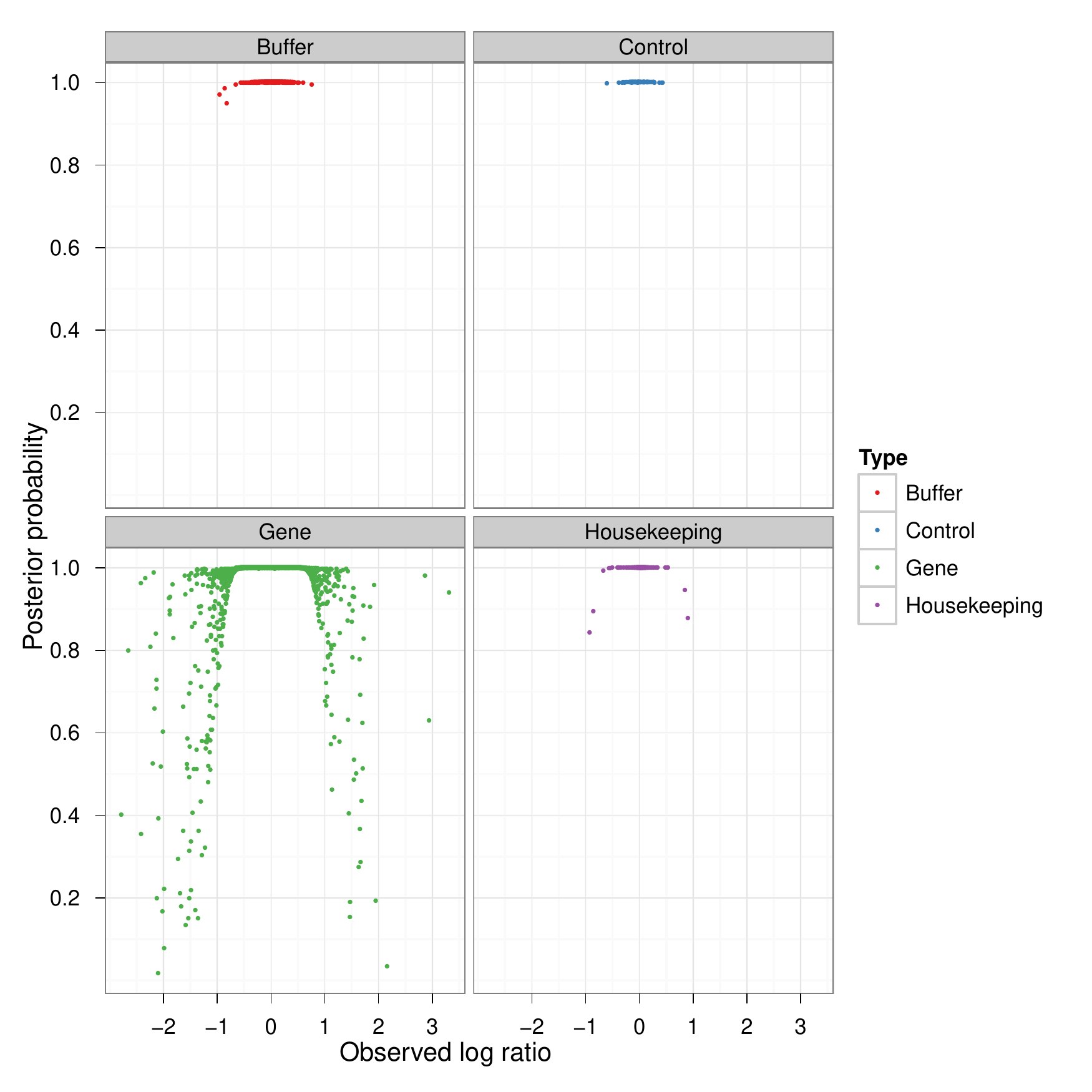}
	\caption{Posterior probability of equivalence versus observed
          mean log ratio for the stem cell genes: day 3 compared to
          day 0.}
	\label{fig:SCPP2}
\end{figure}

A scatter plot of the estimated posterior probability of equivalence
versus the observed mean log ratio is shown in Figure~\ref{fig:SCPP2}
with the genes separated by spot type. The values of the
hyperparameters used to calculate the posterior probabilities are
given in Table~\ref{tab:EM_res} of Appendix 1, and are the same as the values used
for Figure~\ref{fig:EqPP1}.


We note firstly that there are $327$ housekeeping genes spotted on the
microarray, the most common nucleotide being NM\_008084 for which
there are $232$ spots. This nucleotide is a fragment of the gene
glyceraldehyde-3-phosphate dehydrogenase (Gapdh) which has been
validated as a good housekeeping gene in the mouse embryo
\citep{Mamo:2007, Willems:2006}.  Housekeeping genes are highly and
constantly expressed across hybridisations, and are useful for quality
assurance purposes, among other things.  Of the total $327$
housekeeping spots, $316$ ($96.64$\%) have a posterior probability of
equivalence greater than $0.9999$, whilst all $232$ of the NM\_008084
spots have posterior probability of equivalence greater than
$0.9999925$.

The buffer and control genes are expected to have zero gene expression
on both days, and are also equivalently expressed with high posterior
probability, as expected.

Next, we observe some genes with observed mean log ratios lying
outside the equivalence neighbourhood $(-1,1)$, but with posterior
probabilities of equivalence close to one. There are $231$ genes in
total with observed mean log ratios lying outside of the equivalence
neighbourhood, with corresponding posterior probabilities of
equivalence lying between $0.01665$ and $0.99779$.  Of these $231$
genes, $22$ have a posterior probability of equivalence greater than
$0.99$ and all also have a mean log ratio variance greater than
$0.2228$.  This is a large observed variance, as only $2$\% of all
genes on the microarray have an observed variance greater than
$0.2228$, and explains the high posterior probability of equivalence
for these genes, as indicated in Figure 3.  \bigskip


In \cite{Tuke:2009}, the authors identified, with gene profiling, 15
nucleotides whose observed expression levels over time correspond to
the pre-specified profile referred to as the \emph{Oct4} profile. These
nucleotides and associated gene are shown in Table~\ref{tab:oct4pp}
along with their observed posterior probability of equivalence. The
\emph{Oct4} profile requires the nucleotides to have equivalent
expression at day 3 compared to day 0 and so we expected that these
nucleotides would have a posterior probability of equivalence close to
one as seen in the table.

\input{SCPP.tex}

\bigskip

\begin{figure}[htb]
	\begin{center}
          \includegraphics[width=3.5in]{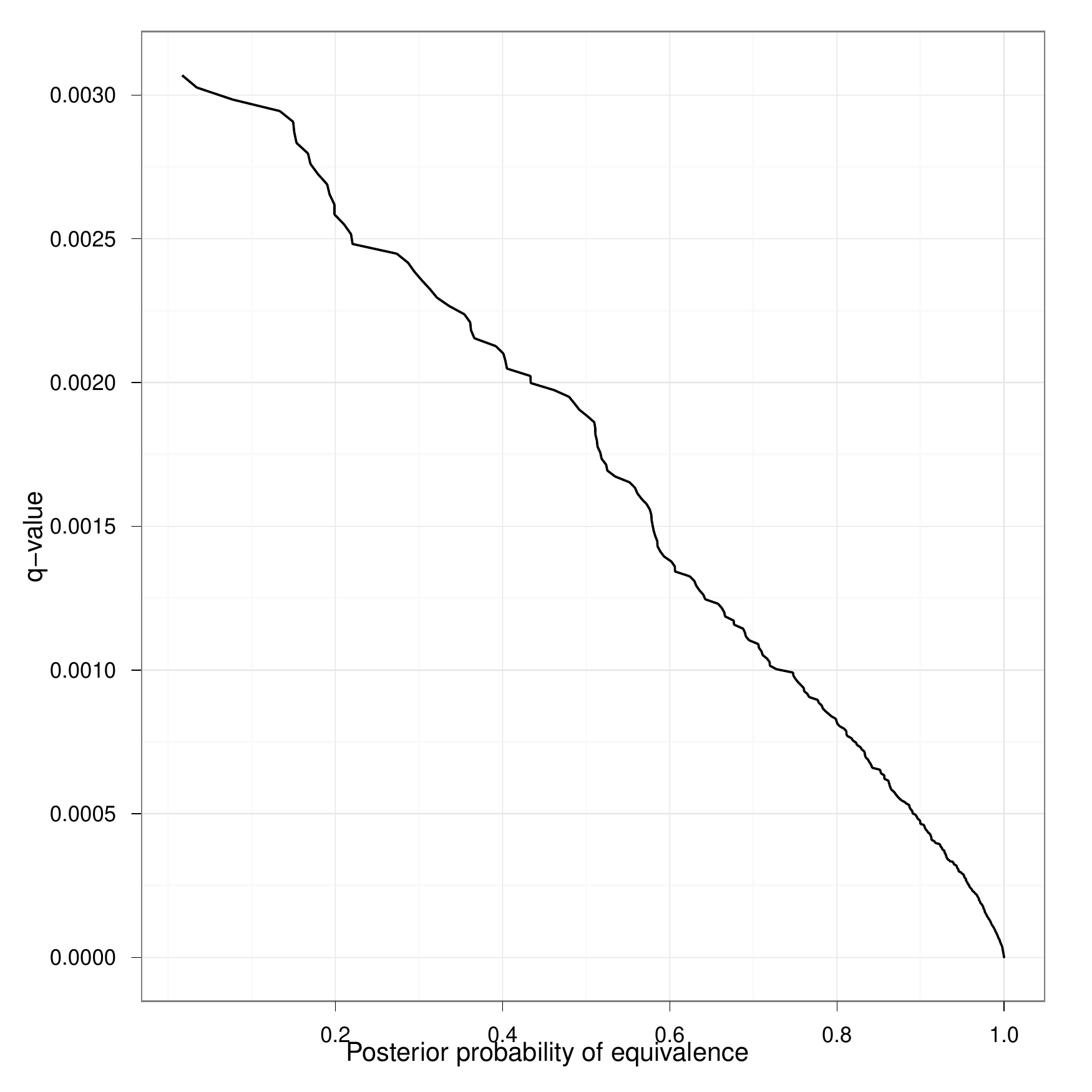}
          \caption{Estimated $q$-value versus posterior probability of
            equivalence for the stem cell data: day 3 compared to day
            0.}
          \label{fig:getSCQv1}
	\end{center}
\end{figure}

Figure~\ref{fig:getSCQv1} plots the estimated $q$-values \eqref{eq:qv} versus
the posterior probabilities of equivalence for the stem cell mean log
ratios comparing day 3 to day 0, and shows good concordance.  We
observe that as the posterior probability of equivalence decreases,
the $q$-value increases to a maximum of 0.003069. This small observed
maximum $q$-value is because the majority of genes are expected to be
equivalently expressed between day 3 and day 0 and so even if all
genes are considered equivalent then the expected proportion of false
positives will be small.
 
\section{Conclusion}\label{sec:con}

We have shown in a general setting that the $q$-value, when
represented as a posterior probability, satisfies some basic
consistency requirements needed for any credible measure of evidence
for equivalence. In particular, the $q$-value increases monotonically
with the variance of the estimator. Importantly, we have established
that the $P$-value does not satisfy these requirements, and hence
should not be used as a measure of evidence for equivalence. Our
results also demonstrate that any attempt to adapt Storey's
\nocite{Storey:2002} approach to estimating $q$-values from
$P$-values in the equivalence testing context is logically untenable,
and propose a simple alternative estimate.  We note that the lack of
duality between $P$-values and $q$-values in equivalence testing
provides some fuel to the debate on the $P$-value as a measure of
evidence {\it per se}, but discussion on this point is beyond the scope of
the present paper.

It is worth remarking here that via simulations conducted (but not
shown), the empirical Bayesian model we assumed of a mixture of three
normal distributions provides a reasonable approximation to the true
posterior probability of equivalence, even in situations where the
true prior distribution of the parameter of interest deviates
considerably from a mixture model of three normals.

\section{Bibliography}
\bibliographystyle{eqpaper}
\bibliography{eqpp}

\section*{Appendix 1: Estimation of  hyperparameters using the EM algorithm}

Applying the EM algorithm to the full probability model \eqref{eq:BM}
for the observed mean log ratios, $Y_i$, we introduce the Bernoulli random
variables $R_{i1}, R_{i2}$ and $R_{i3}$, $i=1,\ldots,23040$, that are
unobserved in the data.

The complete likelihood for the observed data vector $\boldsymbol{Y}$ and the 
unobserved data $\boldsymbol{R_1},\boldsymbol{R_2}$ and $\boldsymbol{R_3}$ is 
then

\begin{equation*}
	L(\boldsymbol{\Psi}|\boldsymbol{Y},
	\boldsymbol{R_{1}},\boldsymbol{R_{2}},\boldsymbol{R_{3}})
	=\prod^{23040}_{i=1}
	\prod^3_{j=1}
	\left[\pi_jg_j(y_i,\mu_j,\tau^2_j)\right]^{R_{ij}},
\end{equation*}
where
\begin{equation*}
	g_j(y_i,\mu_j,\tau^2_j)=\frac{1}{\sqrt{2\pi(\tau^2_j+\sigma^2_i)}}
		\exp\left\{-\frac{(y_i-\mu_j)^2}{2(\tau^2_j+\sigma^2_i)}\right\}
		,j=1,2,3,
\end{equation*}
$\sum_{j=1}^3 \pi_j=1$, and $\boldsymbol{\Psi}=(\pi_1,\pi_2,\pi_3,
\mu_1,\mu_2,\mu_3,
\tau^2_1,\tau^2_2, \tau^2_3)$.

That is, the log-likelihood  is
\begin{align}
	\ell(\boldsymbol{\Psi}|\boldsymbol{Y},
	\boldsymbol{R_1},\boldsymbol{R_2},\boldsymbol{R_{3}})
	=&\sum^{23040}_{i=1}
	\sum_{j=1}^3
	R_{ij}\log\pi_j+R_{ij}
	\log g_i(y_i,\mu_j,\tau^2_j)\label{eq:ELR}
\end{align}

For exponential family distributions, the conditional expectation of the 
complete log likelihood $Q(\boldsymbol{\Psi}|\boldsymbol{\Psi}_n)$, conditional 
on the estimated value $\boldsymbol{\Psi}_n$ of $\boldsymbol{\Psi}$ at the 
$n$th step, can be calculated by substituting 
$E[\boldsymbol{Z}|\boldsymbol{Y},\boldsymbol{\Psi}_n]$ in place of 
$\boldsymbol{Z}$, where $\boldsymbol{Z}$ represents the missing data \citep{Dempster:1977}. For the observed mean log ratio for 
gene $i$,

\begin{align*}
	E[R_{ij}|\boldsymbol{Y},\boldsymbol{\Psi_n}]
		&= P(R_{ij}=1|\boldsymbol{Y},\boldsymbol{\Psi_n})\\
		&=\frac{\pi^{(n)}_j
		g_j(y_i,\mu_j^{(n)},\tau^{2(n)}_j)}
		{\sum_{j=1}^3\pi^{(n)}_j 
		g_j(y_i,\mu_j^{(n)},\tau^{2(n)}_j)},
\end{align*}
where the superscript $(n)$ indicates the parameter estimate from iteration $n$ of the EM algorithm. Substituting $\gamma_{ij} = E[R_{ij}|\boldsymbol{Y},\boldsymbol{\Psi_n}]$ in place of $R_{ij}$ in Equation (\ref{eq:ELR}) gives 
\begin{align}
	Q(\boldsymbol{\Psi}|\boldsymbol{\Psi_n})
	=&E[\log f(\boldsymbol{X}|\boldsymbol{\Psi})
		|\boldsymbol{Y},\boldsymbol{\Psi_n}]\notag\\
	=&\sum^{23040}_{i=1}
	\sum_{j=1}^3\gamma_{ij}\log\pi_j+\gamma_{ij}\log 
	g_j(y_i,\mu_j,\tau^2_j)\label{eq:Q}
\end{align}

In the maximisation step, we use a modification of the EM algorithm
known as the Expectation/Conditional Maximisation (CEM) 
\citep{Meng:1993}. Here, the CEM replaces the M-step with a sequence of
conditional maximisation steps so that the function
$Q(\boldsymbol{\Psi}|\boldsymbol{\Psi_n})$ is maximised for each
parameter in turn, whilst keeping the remaining parameters fixed.
\medskip

Maximisation of $Q(\boldsymbol{\Psi}|\boldsymbol{\Psi_n})$ with
respect to $ \pi_j$, $j=1,2,3$, with the constraint $\sum_{j=1}^3
\pi_j=1$, gives the parameter estimates for $ \pi_j$ at the $(n+1)^{th}$
step as
\begin{align*}
	\hat{\pi}_j^{(n+1)}=
		\frac{\sum^{23040}_{i=1}\gamma_{ij}}
		{\sum^{23040}_{i=1}
		\sum_{j=1}^3\gamma_{ij}},
		\quad j=1,2,3.
\end{align*}

Maximisation of $Q(\boldsymbol{\Psi}|\boldsymbol{\Psi_n})$ with
respect to $ \mu_j$ gives the parameter estimates for $ \mu_j$ at the
$(n+1)^{th}$ step as
\begin{align*}
	\hat{\mu}_j^{(n+1)} = \frac{\sum^{23040}_{i=1}
	\gamma_{ij}y_i/(\sigma^2_i+\hat{\tau}^{2(n)}_j)}
	{\sum^{23040}_{i=1}
	\gamma_{ij}/(\sigma^2_i+\hat{\tau}^{2(n)}_j)},
	\quad j=1,2,3.
\end{align*}

Note the use of the estimates from the previous M-step for $\tau^2_j$, $j=1,2,3$. 
\bigskip

\noindent {\it Numerical methods}: We used numerical optimisation, in
particular, the {\tt R} \citep{R-Development-Core-Team:2011} function
{\tt optimise} \citep{Brent:2002} to find the values of $\tau^2_j$
that maximise
$\sum_{i=1}^{23040}\gamma_{ij}g_j(y_i,\mu^{(n+1)}_j,\tau^2_j)$
for $ j=1,2,3$. The {\tt R} function {\tt optimise} uses a combination
of golden section search and successive parabolic interpolation to
find the maximum over an interval. We considered the interval $0\leq
\tau^2_j \leq \max( \sigma^2_i)$, $ i=1,\ldots,23040; j=1,2,3$. Note
again the use of the $(n+1)^{th}$ parameter estimate for $\mu_j$.
\medskip

\noindent {\it Choosing initial parameter values}: We modified a method 
proposed by Finch \emph{et al.} to obtain initial estimates of the 
hyperparameters given an initial choice of the mixing proportions $\pi_j$, 
$j=1,2,3$ for the EM algorithm. For each gene, there are two observations: the 
mean log ratio $y_i$ and the variance $ \sigma^2_i$. These pairs of 
observations were ordered according to the mean log ratios $y_i$. These values 
were then split into three samples consisting of the smallest $n_1$ 
observations, the smallest $n_2$ of the remaining observations, and the 
remaining observations, where $n_1$ is the value $23040\pi_1$ rounded to the 
nearest integer and $n_2$ is the value $23040 \pi_2$ rounded to the nearest 
integer. For each of these three samples, the estimate of $ \mu_j$ was 
initialised as the sample mean of the observed mean log ratios. The initial 
value of $ \tau^2_j$ was obtained by finding the value $\tau^2_j$ that maximises

\[
\sum_{i=1}^{n_j}\log\left[
	\frac{1}{\sqrt{2\pi(\tau^2_j+\sigma^2_i)}}
	\exp\left\{-\frac
	{(y_i-\hat \mu_j)^2}
	{2(\tau^2_j+\sigma^2_i)}\right\}\right],		
\]
where $n_j$ is the number of observations in the $j$th sample, $y_i$
and $\sigma^2_i$ are the pair of observations for each gene in the
$j$th sample, and $\hat \mu_j$ is the sample mean of the $y_i$ in the
$j$th sample. The {\tt R} function {\tt optimise} was used to find
this maximum over the range $0\leq \tau^2_j \leq \max( \sigma^2_i),
i=1,\ldots,n_j$, $ j=1,2,3$.

\cite{Karlis:2003} compare via simulations a number of methods for
choosing initial values for the EM algorithm, including the one
proposed by \cite{Finch:1989}.  They observed that for two-component
and three-component mixtures with equal mixing proportions for the
initial values, \nocite{Finch:1989} Finch \emph{et al.}'s method
generally locates the \lq global maximum' more often than other
methods considered.  They recommend that a mixed strategy is used in
the choice of initial values in the EM algorithm, as different initial
values may find a local but not a global maximum.  

For each choice of initial parameter values, the EM algorithm is
iterated for a limited number of steps. The EM algorithm is then run
with the initial parameter values that give the largest likelihood for
the initial iterations. In this final iteration, the EM algorithm is
iterated until a high level of accuracy in the parameter estimates is
achieved.

\medskip

For the stem cell data, the initial values for the seven main mixing
proportions considered are set out in Table~\ref{tab:EMP2}. These are: equally
contributing normals (Run 1); a single dominating normal (Runs 2, 3
and 4); and two dominating normals (Runs 5, 6 and 7).  For each choice
of the initial parameters, the EM algorithm was run for a limited
number of steps, and the choice of initial parameters which produced
the maximum log-likelihood was then repeated for a larger number of
iterations.

\input{EM_results2.tex}

After each iteration of the EM algorithm, the absolute difference in
each of the parameter estimates compared to the previous iteration was
calculated, and the iterations continued until the maximum of these
differences was less than $10^{-5}$.
The top five runs ordered by log-likelihood showed consistent
parameter estimates, and the initial values, $ \pi_1=0.45, \pi_2=0.1,
\pi_3=0.1$, produced the maximum log-likelihood of $5368.77$.  These
estimates were then used for the initial values of the EM algorithm
which was repeated until the maximum difference in consecutive
parameter estimates was less than $10^{-10}$.

The resulting final parameter estimates for the stem cell data are
given in Table~\ref{tab:EM_res}.

\bigskip

\input{EM_results.tex}

\section*{Appendix 2: Monotonicity of posterior probabilities of equivalence}

\noindent {\bf Lemma 1}: Suppose \(f(x)\) is a symmetric function with 
\(f(x)>0\) for all \(x\) and consider numbers  \(a,b,c,d,\ell\) with
\(-\ell<a<b<\ell\), \(0<c\), \(\ell<d\) and \(b-a=d-c\). Then
\[
\frac{\int_a^bx^2f(x)dx}{\int_a^bf(x)dx}
<
\frac{\int_c^dx^2f(x)dx}{\int_c^df(x)dx}
\]
\medskip

\begin{proof}
By symmetry, we can assume \(b>0\) and \(|a|\leq |b|\).
It is then sufficient to consider the following three cases.

\begin{enumerate}
\item{Case 1:}{\(b\leq c\):
Observe that 
\[
\frac{\int_a^bx^2f(x)dx}{\int_a^bf(x)dx}
<b^2
\mbox{ and }
\frac{\int_c^dx^2f(x)dx}{\int_c^df(x)dx}>c^2
\]
so the result follows.}

\item{Case 2:}{\(c<b\) and \(|a|\leq c\):\newline
Let
\[
m_1=\frac{\int_a^cx^2f(x)dx}{\int_a^cf(x)dx},\
m_2=\frac{\int_c^bx^2f(x)dx}{\int_c^bf(x)dx}
\mbox{ and }
m_3=\frac{\int_b^d x^2f(x)dx}{\int_b^df(x)dx}
\]
and observe that
\(m_1<c^2<m_2<b^2<m_3\).
Since
\[
\frac{\int_a^bx^2f(x)dx}{\int_a^bf(x)dx}
\]
is a weighted average of \(m_1\) and \(m_2\) and 
\[
\frac{\int_c^dx^2f(x)dx}{\int_d^df(x)dx}
\]
is a weighted average of \(m_2\) and \(m_3\),
the result follows.}
\item{Case 3:}{ \(c<b\) and \(|a|>c\):\newline
To have \(|a|>c\) requires \(a\) to be negative. Now let
\begin{eqnarray*}
m_1= \frac{\int_a^{-c}x^2f(x)dx}{\int_a^{-c}f(x)dx},\
m_0=\frac{\int_{-c}^cx^2f(x)dx}{\int_{-c}^cf(x)dx},\
m_2=\frac{\int_c^bx^2f(x)dx}{\int_c^bf(x)dx}\\
\mbox{ and }\
m_3=\frac{\int_b^d x^2f(x)dx}{\int_b^df(x)dx},
\end{eqnarray*}
and observe \(m_0<c^2<m_1<m_2<b^2<m_3\).
Since
\[
\frac{\int_a^bx^2f(x)dx}{\int_a^bf(x)dx}
\]
is a weighted average of \(m_0,\ m_1\) and \(m_2\) and 
\[
\frac{\int_c^dx^2f(x)dx}{\int_d^df(x)dx}
\]
is a weighted average of \(m_2\) and \(m_3\),
the result follows.}
\end{enumerate}
\end{proof}

\noindent {\bf Theorem 2}: Suppose \(\theta\) has prior distribution \(P(\theta)\) and 
\(T|\theta\sim N(\theta,\sigma^2)\). Consider numbers \(0<\ell<\varepsilon\)
and assume further that \(0<P(-\varepsilon<\theta<\varepsilon)<1\).
Then \(P(-\varepsilon<\theta<\varepsilon\mid-\ell<T<\ell)\) is a 
decreasing function of \(\sigma^2\).
\medskip

\begin{proof}
Observe that 
\begin{eqnarray*}
P(-\varepsilon<\theta<\varepsilon\mid-\ell<T<\ell)
&=&
\frac
{\int_{|\theta|\leq\varepsilon} P(\theta)\int_{-\ell}^\ell 
e^{-\frac{1}{2\sigma^2}(t-\theta)^2}dtd\theta}
{\int_{-\infty}^\infty P(\theta)\int_{-\ell}^\ell 
e^{-\frac{1}{2\sigma^2}(t-\theta)^2}dtd\theta}\\
&=&
\frac
{\int_{|\theta|\leq \varepsilon} P(\theta)\int_{-\ell-\theta}^{\ell-\theta} 
e^{-\frac{1}{2\sigma^2}y^2}dyd\theta}
{\int_{\theta\leq \varepsilon} P(\theta)\int_{-\ell-\theta}^{\ell-\theta} 
e^{-\frac{1}{2\sigma^2}y^2}dyd\theta+
\int_{|\theta|>\varepsilon} P(\theta)\int_{-\ell-\theta}^{\ell-\theta} 
e^{-\frac{1}{2\sigma^2}y^2}dyd\theta}.
\end{eqnarray*}
Taking \(\omega=1/(2\sigma^2)\), it is sufficient to show that the 
posterior odds
\[
r(\omega)= \frac
{\int_{|\theta|\leq\varepsilon} P(\theta)\int_{-\ell-\theta}^{\ell-\theta} 
e^{-\omega y^2}dyd\theta}
{\int_{|\theta|>\varepsilon} P(\theta)\int_{-\ell-\theta}^{\ell-\theta} 
e^{-\omega y^2}dyd\theta}.
\]
is increasing in \(\omega\).
Now observe
\begin{eqnarray*}
r'(\omega)&=&
\left\{
\left(\int_{|\theta|\leq\varepsilon}P(\theta)\int_{-\ell-\theta}^{\ell-\theta} 
e^{-\omega y^2}dyd\theta\right)
\left(\int_{|\theta|>\varepsilon} P(\theta)\int_{-\ell-\theta}^{\ell-\theta} 
y^2e^{-\omega y^2}dyd\theta\right)\right.\\
&&-\left.
\left(\int_{|\theta|\leq\varepsilon} P(\theta)\int_{-\ell-\theta}^{\ell-\theta} 
y^2e^{-\omega y^2}dyd\theta\right)
\left(\int_{|\theta|>\varepsilon} P(\theta)\int_{-\ell-\theta}^{\ell-\theta} 
e^{-\omega y^2}dyd\theta\right)\right\}\\
&&\bigg/
\left\{
{\int_{|\theta|>\varepsilon} P(\theta)\int_{-\ell-\theta}^{\ell-\theta} 
e^{-\omega y^2}dyd\theta}\right\}^2.
\end{eqnarray*}
Taking
\[
A_1=\inf_{|\theta|>\epsilon}\frac
{\int_{-\ell-\theta}^{\ell-\theta}y^2e^{-\omega y^2}dy}
{\int_{-\ell-\theta}^{\ell-\theta} e^{-\omega y^2}dy}
\mbox{ and }
A_0=\sup_{|\theta|\leq\epsilon}\frac
{\int_{-\ell-\theta}^{\ell-\theta}y^2e^{-\omega y^2}dy}
{\int_{-\ell-\theta}^{\ell-\theta} e^{-\omega y^2}dy},
\]
it follows that 
\[
r'(\omega)\geq
(A_1-A_0)\frac{
\left(\int_{|\theta|\leq\varepsilon}P(\theta)\int_{-\ell-\theta}^{\ell-\theta} 
e^{-\omega y^2}dyd\theta\right)\left(
{\int_{|\theta|>\varepsilon} P(\theta)\int_{-\ell-\theta}^{\ell-\theta} 
e^{-\omega y^2}dyd\theta}\right)}
{
\left\{
{\int_{|\theta|>\varepsilon} P(\theta)\int_{-\ell-\theta}^{\ell-\theta} 
e^{-\omega y^2}dyd\theta}\right\}^2}
\]
By Lemma 1, \(A_1>A_0\) so the proof is complete.
\end{proof}
\end{document}

%% file: SCPP.tex
\begin{table}[htbp]
\begin{center}
\begin{tabular}{l|r}
  \hline
Nucleotide & Posterior Probability \\ 
  \hline
NM\_013633 (Oct4) & 0.998 \\ 
  NM\_009482 (Utfl) & 0.9989 \\ 
  NM\_011562 (Tdgf1) & 0.9983 \\ 
  AK005182 (Slc35f2) & 0.9992 \\ 
  NM\_009426 (Trh) & 0.9985 \\ 
  NM\_010425 (Foxd3) & 0.9993 \\ 
  AF246632 (Musd1) & 0.9988 \\ 
  BC004805 (Skil) & 0.9986 \\ 
  NM\_010127 (Pou6f1) & 0.9988 \\ 
  NM\_007974 (Par2) & 0.9974 \\ 
  AK010332 (Nanog) & 0.9966 \\ 
  NM\_007515 (Slc7a3) & 0.9986 \\ 
  NM\_010316 (Gng3) & 0.9994 \\ 
  NM\_011386 (Skil) & 0.9986 \\ 
  NM\_007905 (Rae-28) & 0.9996 \\ 
   \hline
\end{tabular}
\caption{Nucleotides identified by gene profiling with the \emph{Oct4} profile \citep{Tuke:2009} and their associated posterior probability of equivalence.}
\label{tab:oct4pp}
\end{center}
\end{table}

%% file: EM_results2.tex
\begin{table}[ht]
\begin{center}
\begin{tabular}{l|rrr}
  \hline
Run & Initial $\pi_1$ & Initial $\pi_2$ & Initial $\pi_3$ \\ 
  \hline
  1 & 0.33 & 0.33 & 0.33 \\ 
    2 & 0.8 & 0.1 & 0.1 \\ 
    3 & 0.1 & 0.8 & 0.1 \\ 
    4 & 0.1 & 0.1 & 0.8 \\ 
    5 & 0.1 & 0.45 & 0.45 \\ 
    6 & 0.45 & 0.1 & 0.45 \\ 
    7 & 0.45 & 0.45 & 0.1 \\ 
   \hline
\end{tabular}
\caption{Initial mixing proportion for EM algorithm.}
\label{tab:EMP2}
\end{center}
\end{table}

%% file: EM_results.tex
\begin{table}[htbp]
\begin{center}
\begin{tabular}{l|r}
  \hline
Parameter & Estimate \\ 
  \hline
$\pi_1$ & 0.03177 \\ 
  $\pi_2$ & 0.3576 \\ 
  $\pi_3$ & 0.6107 \\ 
  $\mu_1$ & -0.09135 \\ 
  $\mu_2$ & -0.01845 \\ 
  $\mu_3$ & 0.008169 \\ 
  $\tau_1^2$ & 0.3558 \\ 
  $\tau_2^2$ & 0.01958 \\ 
  $\tau_3^2$ & 5.426e-12 \\ 
   \hline
\end{tabular}
\caption{Parameter estimates from the EM algorithm for stem cell data.}
\label{tab:EM_res}
\end{center}
\end{table}